\documentclass[aps,pra,showpacs,twoside,twocolumn,10pt]{revtex4-1}
\usepackage[colorlinks=true, citecolor=red, urlcolor=blue ]{hyperref}
\usepackage{times,epsfig,amssymb,amsfonts,amsmath,bm,subfigure,mathtools,amsthm,braket,soul,enumitem,color,physics,graphics,graphicx}
\usepackage[normalem]{ulem}
\usepackage{comment}
\usepackage{epstopdf}

\begin{document}

\title{
Manifestation of Rank-Tuned Weak Measurements Towards Featured State Generation
}

\author{Pritam Haldar$^{1}$, Ratul Banerjee$^{1}$, Shiladitya Mal$^{2, 3}$, Aditi Sen(De)$^{1}$}

\affiliation{$^1$ Harish-Chandra Research Institute, A CI of Homi Bhabha National
Institute,  Chhatnag Road, Jhunsi, Allahabad - 211019, India\\
		$^2$ Physics Division, National Center for Theoretical Sciences, Taipei 10617, Taiwan \\
$^3$ Department of Physics and Center for Quantum Frontiers of Research and Technology (QFort),
National Cheng Kung University, Tainan 701, Taiwan	
	}

\begin{abstract}
 We propose that an unsharp measurement-based process to generate genuine multipartite entanglement from an entangled initial state with a fewer number of qubits can be classified in two ways -- biased and unbiased inflation protocols. In the  biased case, genuine multipartite entanglement (GME) of the resulting state obtained after a single measurement outcome is optimized, thereby creating a possibility of states with high GME while in the unbiased case, average GME is optimized over all possible outcomes. Interestingly, we show that the set of two-qubit unsharp measurements can generate multipartite states having different features according to GME measure, generalized geometric measure,  the monogamy-based entanglement measure, tangle and robustness against particle loss quantified via persistency depending on the rank of the unsharp measurement operators. Specifically, in the process of producing three-qubit pure states, we prove that rank-$2$ measurements can create only Greenberger Horne Zeilinger (GHZ)-class states while only W-class states are produced with rank-$4$ measurements although rank-$3$ measurements are capable to generate both. In the case of multipartite states with an arbitrary number of qubits, we report that the average content of genuine multipartite entanglement increases with the decrease of the rank in the measurement operators although the persistency decreases with the rank, both in the biased as well as unbiased protocols.

\end{abstract}	

\maketitle

\section{Introduction}

The peculiarities  of quantum physics are revealed by different non-classical resources \cite{horodecki09,modi12, stretslov17, bera18} which have no classical counterpart, and are useful for a variety of  quantum information processing tasks \cite{Kozlowski2019,Azuma2021,Pirker2018,Pirker2019,Meignant2019,Gyongyosi2019,Miguel2020,Miguel2021} such as quantum communication \cite{tele-pan-photon,tele@bennett,dc,DC_obstacle,dcrev} including quantum key distribution \cite{crypto1,BB84,crypto2,Crypto3,Crypto4,Keydist}, quantum metrology \cite{Giovanneti2011} and distributed quantum computation \cite{cirac99}.
Significant quantum properties that enable nonclassical efficiency include quantum coherence  \cite{stretslov17}, bipartite as well as multipartite entanglement \cite{horodecki09}, correlations beyond entanglement \cite{modi12, bera18}. In order to classify resource states, which are the fundamental components of quantum technologies, and to explain quantum foundational problems, this is an attractive path to explore.



Building both quantum networks and quantum computation has necessitated the development of preparation processes for multipartite entangled states, and it becomes of utmost importance in recent years. For instance, a viable alternate route to the circuit-based strategy \cite{nielsen_chuang_2010} for achieving practical quantum information processing has been made possible by measurement-based quantum computing \cite{Rausendorf2001, Briegel2009}, which critically depends on a multipartite entangled state known as a cluster state \cite{persistency01} and local projective measurements. Thus, it is vital to look into how various measurement strategies can result in different resource states in the context of quantum computation and other quantum protocols.


Information gain can be at its highest under the von-Neumann or standard projective measurement scenario \cite{nielsen_chuang_2010}, although they also disturb the system. However, a weak or unsharp measurement \cite{busch86,busch96} can take advantage of the trade-off between information gain and disturbance so that it disturbs the state only minimally while obtaining the data that is required. 
Notice that we deal with the weak (unsharp) measurements without taking weak values in account \cite{Aharonov1988}. 
Numerous scenarios have demonstrated the benefits of weak measurements, including joint measurability, state discrimination, tomography of states, violation of Bell inequalities, randomness generation \cite{derka98,shang18,dieks88,peres88,vertesi10,gomez18}, sequential sharing of non-locality  \cite{silva15,Mal16}, binonlocality \cite{halder2022limits, bera18}, and teleportation or telecloning fidelity \cite{ROY2021127143,das2022sequential} etc to name a few.

Based on weak measurements,
we have recently proposed a scheme, called ``measurement-based multipartite entanglement inflation'' \cite{pritam_inflation_21} to generate genuine multipartite entangled states.
 In order to create an entangled state with a bigger number of parties — which is not achievable with sharp entangling measurement — weak entangling measurement is applied to one of the parties of the initial entangled resource of less number of sites and another to an auxiliary state as fundamental units of operation. 
In this study, we demonstrate how the different classes of multipartite 
entangled states are produced by tuning the measurement parameters. We also investigate the relationship between the type of unsharp measurements  and the multiparty entanglement of the created state. 
We choose the rank  to characterize  weak measurement operators which are  same for all the operators and  different multipartite entanglement quantifiers are adopted to characterize  multiparty states.

Precisely, we will address the following questions.
\emph{Is it possible to create different classes of quantum states via unsharp measurements of various ranks? If so, is there a connection between entanglement properties or any other features of the resulting state, and the rank of the unsharp measurement operators? }
Our work provides affirmative answers to both of these questions. We indeed determine that if the feature of unsharp measurement is taken to be the rank of the measurement operator, we can classify the generated states in the entanglement inflation process. The classification is made by computing their genuine multipartite entanglement content, quantified by generalized geometric measure (GGM) \cite{aditi2010}, quantum correlation measure based on monogamy inequality \cite{coffman00}, tangle \cite{coffman00,dur00}, and robustness against particle loss, called persistency \cite{persistency01}. If a genuine multipartite entanglement of the resulting state is maximized over a specific measurement outcome, we call it as a biased inflation process while when the average multipartite entanglement content of the output states is optimized over all measurement outcomes, we refer to it as the unbiased ones.

In particular, it is already known that two inequivalent classes of entangled states via stochastic local operations and classical communication (SLOCC) exist for three-qubit pure states, the so-called Greenberger Horne Zelinger- \cite{Greenberger2007} and the W-class states \cite{Durvidal2002}. When compared to all other pure states, the W-class states constitute a set of measure zero, and an entanglement monotone known as tangle vanishes  \cite{coffman00,dur00} for them while remaining non-vanishing for the GHZ-class states. We demonstrate a weak measurement-based approach that, by varying the rank of measurement, can distinguish between several kinds of three-qubit states. However, states with more than three parties can have an infinite number of inequivalent classes  under SLOCC, making it difficult to generalise the approach used for three qubits.
However, we illustrate that persistency (which is defined as a minimum number of single-qubit local measurements applied to disentangle the state) \cite{persistency01},  tangle \cite{dur00} and GGM together can be used to categorise multipartite states with more than three parties produced by the varying rank of the unsharp measurement operators. We report that both in the biased and unbiased entanglement inflation protocols, high genuine multipartite entangled state  can be produced with Haar uniformly generated initial resource states and a suitable auxiliary state although the entanglement content of the resulting state decreases on average with the increase of the rank of the unsharp measurement.


The  remainder of our paper has the following structure. In Sec. \ref{sec:povm}, we discuss the construction of weak measurements of various ranks used throughout this work. Sec. \ref{sec:protocol} illustrates the state generation protocol and its classification. Discrimination protocol of three-qubit GHZ- and the W-class states are presented in Sec. \ref{sec:3qubit} based on the rank of the measurement operators while the classification of multiparty entangled states (more than three parties) are described in Sec. \ref{sec:multiqubit}. A similar classification of multiparty entangled states with the help of unbiased protocol is reported in Sec.\ref{subsec:unbiased state gen}.  Finally, Sec. \ref{subsec:optimal_resource_distribution} deals with the role of the optimal resource state in the creation of a multipartite entangled state and the concluding remarks are added in Sec. \ref{sec:conclusion}.

\section{Introducing Unsharp measurements of different ranks}
\label{sec:povm}

The main goal of this work is to create genuine multiparty entangled (GME) states, which may be suitable for quantum  information processing tasks. We attempt a measurement-based approach to achieve the same. 

The most general quantum measurement is described by a set of $n$-outcome positive operators, $\{M_k\}_{k=1}^n$, which satisfy $\sum_{k}M_k=\mathbb{I}$, known as positive operator valued measure (POVM). Notice that the specific class of measurement, called projective (sharp) measurements (PV), when applied to a  specific subsystem of a multiqubit state, gets disentangled from the rest of the system which is not the case for POVMs. 
 Moreover, for a given dimension, the number of elements in the set of rank-$1$ PV measurements is equal to the dimension of the system and they form an orthogonal basis. In particular,  PVs can have $d$ outcomes, where $d$ signifies the dimension of the Hilbert space in which measurement is performed while 
it is sufficient for a POVM having  $d^2$ outcomes. 
  
Our idea is to employ POVM operators to
  convert entangled states with fewer qubits into multipartite entangled states by keeping the number of outcomes in POVM  bounded by the dimension.
  Specifically, let us consider a projective measurement, $\{M_i=\ket{\Phi_i}\bra{\Phi_i}\}_{i=1}^d$, where $\{\ket{\Phi_i}\}_{i=1}^d$ forms a basis in the $d$ dimensional Hilbert space. By introducing the noise parameter $p \in [0,1]$, we can design the unsharp measurement originated from the sharp ones as $\big\{M_k = \sum_{i=1}^d p_i\ket{\Phi_i}\bra{\Phi_i}\big\}_{k=1}^n$ with the condition $\sum_{i=1}^d p_i = 1$ and $2 \leq n \leq d$. The number of elements in the measurement operator, $M_k$, i.e., the maximum value of \(i\) represents the rank of the  operator. Moreover, in this work, we assume that the set of unsharp measurement  consists of measurement operators having same ranks. We will call the rank of the individual operators as the rank of the unsharp measurements. 

\textit{An example of a class of unsharp measurements.} To illustrate about the construction of the unsharp measurement, let us consider a two-qubit PV measurement in four dimension, namely Bell basis $\{|\phi^{\pm}\rangle = \frac{1}{\sqrt{2}}(|00\rangle \pm |11\rangle), |\psi^{\pm}\rangle = \frac{1}{\sqrt{2}}(|01\rangle \pm |10\rangle)\}$. It can be represented as 
\begin{eqnarray}
\nonumber \widetilde{M_1} &=& \ket{\phi^+}\bra{\phi^+} ;
\widetilde{M_2} = \ket{\phi^-}\bra{\phi^-},\\\widetilde{M_3} &=& \ket{\psi^+}\bra{\psi^+};
\widetilde{M_4} = \ket{\psi^-}\bra{\psi^-}.
\end{eqnarray}
A possible example of unsharp measurement operator having rank-$4$ can be obtained by taking admixture of Bell state with white noise, i.e., $\bigg\{M_k = p\widetilde{M_k}+(1-p)\frac{\mathbb{I}}{4}\bigg\}_{k=1}^4$ \cite{pritam_inflation_21,Mal16,halder2022limits,das2022sequential} where $p$ is the control parameter. Instead of white noise, the other set of unsharp measurements can be designed by mixing a particular Bell state with other Bell states, which can be interpreted as a mixture of Bell state with coloured noise. This method can lead to the four-outcome unsharp measurements having different ranks. Specifically, the $d$-outcome weak measurement of different ranks $r$ ($\geq 2$) can be represented as
\begin{eqnarray}
\bigg\{ M_k^r = \sum_{i=k}^{k+r-1}p_i \widetilde{M_i} \bigg\}_{k=1}^d, \hspace{0.2cm} \widetilde{M}_{d+i} = {\widetilde{M_{i}}},
\label{eq:rpovm}
\end{eqnarray}
where, for example, the mixing of probability can be chosen as 
\begin{eqnarray*}
&p_i& = 
        \begin{cases}
            p & \text{if $k\leq i<(k+r-1)$},\\\nonumber
            1-(r-1)p & \text{if $i=k+r-1$},  
        \end{cases}
\end{eqnarray*}
and  $0\leq p < \frac{1}{r-1}$. Based on the Bell basis, let us write unsharp measurements with rank-$2$, -$3$, -$4$ explicitly as follows:\\
For $r=2$, it reads as 
\begin{eqnarray}
\nonumber \big\{M_1^{2} &=& p\ket{\phi^+}\bra{\phi^+}+(1-p)\ket{\phi^-}\bra{\phi^-},\\
\nonumber M_2^{2} &=& p\ket{\phi^-}\bra{\phi^-}+(1-p)\ket{\psi^+}\bra{\psi^+},\\
\nonumber M_3^{2} &=& p\ket{\psi^+}\bra{\psi^+}+(1-p)\ket{\psi^-}\bra{\psi^-},\\
M_4^{2} &=& p\ket{\psi^-}\bra{\psi^-}+(1-p)\ket{\phi^+}\bra{\phi^+}\big\},
\label{eq:rank2povm}
\end{eqnarray}
while the first element of rank-$3$ and rank-$4$ unsharp measurements can be written respectively as
\begin{eqnarray}
\nonumber
M_1^{3} = p\ket{\phi^+}\bra{\phi^+}+p\ket{\phi^-}\bra{\phi^-} 
+ (1-2p)\ket{\psi^+}\bra{\psi^+},\\
\label{eq:rank3povm}
\end{eqnarray}
and
\begin{eqnarray}
\nonumber
M_1^{4} &=& p\ket{\phi^+}\bra{\phi^+} + p\ket{\phi^-}\bra{\phi^-} + p\ket{\psi^+}\bra{\psi^+}\\
    &+& (1-3p)\ket{\psi^-}\bra{\psi^-}.
\label{eq:rank4povm}
\end{eqnarray}
Note that the rank-$4$ measurements can also be obtained by mixing Bell states with white noise. Keeping the constraint $\sum_{k=1}^d M_k^r=\mathbb{I}$ and following Eq. (\ref{eq:rpovm}), one can easily derive rest of the elements in the set. Notice that the structure of unsharp measurements based on a single basis, e.g., the Bell basis is also not unique and the infinite number of such sets can be generated by different choices of $p_i$ in Eq. (\ref{eq:rpovm}). We will manifest in the succeeding section that different classes of GME states having distinct features can be produced according to the rank of the unsharp measurement operators  constructed above.

\section{Unsharp Measurement-based state generation and its classification}
\label{sec:protocol}

Recently, we have introduced a protocol based on unsharp measurements, called entanglement inflation method, to create  GME states from an initial resource consisting of entangled states with fewer number of parties and a single-qubit auxiliary system \cite{pritam_inflation_21}. Based on this work, we show here that  a more general framework for preparing GME states  can be developed. It can be of two types - $(i)$ \emph{biased entanglement inflation} and $(ii)$ \emph{ unbiased entanglement inflation}. Let us describe these state generating protocols in details.



\subsection{ Biased entanglement  inflation}
The biased  inflation protocol is based on the optimization of a property of the output state obtained via a single outcome of a measurement and hence the name. The steps of the procedure for three-qubits can be described as follows:
\begin{enumerate}
    \item Initially, we take a two-qubit entangled state and a single-qubit auxiliary state in a product form, $\ket{\psi_{in}} \otimes \ket{\psi_{aux}}$, where $\ket{\psi_{aux}}= \alpha\ket{0}+\beta\ket{1}$ with $\alpha=\cos \frac{\theta_a}{2}, \beta = e^{i\phi_a}\sin \frac{\theta_a}{2}$.
    
    \item Unsharp measurements of different ranks  are applied on one of the qubits of $\ket{\psi_{in}}$ and $\ket{\psi_{aux}}$ while other qubits remain untouched, to create a genuine three-qubit entangled state. When outcome $k$ clicks, we obtain
    \begin{eqnarray}
    \ket{\psi_{out,k}^r}_{3} = \frac{\sqrt{M_{k}^r(p)}\big(\ket{\psi_{in}}\otimes\ket{\psi_{aux}}\big)}{\sqrt{q_k}}
    \end{eqnarray}
    where $q_{k} = \bra{\psi_{aux}}\otimes \bra{\psi_{in}}M_k^r(p)\ket{\psi_{in}}\otimes\ket{\psi_{aux}}$ is the probability of obtaining  outcome $k$ of the measurement.
    
    \item The optimal output state is created after optimizing genuine multipartite entanglement measure,  $(\mathcal{Q})$, over a single outcome of all unsharp measurements, and auxiliary state parameters for a fixed initial state, i.e., 
    \begin{eqnarray}
    \mathcal{Q}^B_{\max} = \underset{{M_k^r(p),\theta_a, \phi_a}}{\max} \hspace{0.2cm} \mathcal{Q} ( \ket{\psi_{out,k}^r}_3)
      \label{protocol1}
    \end{eqnarray}
     where $M_k^r(p)$, and  $\{\theta_a, \phi_a\} $ are unsharp measurement operators and the state parameters respectively. Suppose the maximization is achieved for the outcome $k$ at $M^{'r}_k(p^{cr}),\theta_a^{cr}$, and $\phi_a^{cr}$ and hence all the corresponding output states for different outcomes are calculated using these choices of parameters. Notice that the multipartite entanglement content can typically be lower for other measurement outcome $M_j^r(p)(j \neq k)$, than the outcome, $k$ for which the optimization is carried out. 
     
     After creating a three-qubit state, similar steps can be repeated to produce GME states with more number of parties. Specifically, a two-qubit entangled state and $(N-2)$ auxiliary qubits, along with repeated unsharp measurements can lead to $N$-qubit GME state. 
\end{enumerate}

\subsection{Unbiased entanglement inflation}

The protocol in this case remains same as the biased one except the optimization process. Instead of optimizing over a single outcome, in this scenario,  the average genuine multiparty entanglement content, $\mathcal{Q}$ of the output state over different outcomes is maximized over the auxiliary state parameters and unsharp measurements. Therefore, we can write
\begin{eqnarray}
      \mathcal{Q}^{UB}_{\max} = \underset{{M_k^r(p),\theta_a, \phi_a}}{\max} \hspace{0.2cm}\sum_{k=1}^{4} q_{k} \mathcal{Q} (\ket{\psi_{out,k}^r}_3).
      \label{protocol2}
\end{eqnarray}
The resulting states after the process are obtained by using the optimal choices of auxiliary states and measurements which are same for all the outcomes. 

\textbf{Remark 1.} In this work, we will use generalized geometric measure \cite{aditi2010} as genuine multipartite entanglement measure. 
A pure state is said to be  genuinely multipartite entangled when it is not product across any bipartitions. A distance-based measure, GGM, of a multipartite state, \(\ket{\Psi}\) is  defined as the minimum distance between a nongenuinely multipartite entangled state and a given state. Mathematically, 
\(\mathcal{G} = 1- \max |\langle \phi | \Psi\rangle|^2\), where maximization is performed over the set of nongenuinely multipartite entangled states. Interestingly, it can be shown that \(\mathcal{G}\) can be calculated by using Schmidt coefficients in different bipartitions of \(\ket{\Psi}\).

\textbf{Reamark 2.} We will exhibit that  both the methods described above are capable to produce   different classes of multipartite states by adjusting the rank of unsharp measurements  appropriately. We will discriminate different types of states with the help of entanglement measures like GGM, tangle \cite{coffman00,dur00}  and persistency \cite{persistency01}, a measure for robustness of entanglement against particle loss.

\section{Classifying three-qubit states via rank of measurements}
\label{sec:3qubit}

Our aim is to show here that the states with contrasting features can be generated when unsharp measurement operators having various ranks are applied. Before investigating it, let us discuss the known classes in pure three-qubit states. Two states are said to be inequivalent via stochastic local operations and classical communication (SLOCC)  if one can not be transformed to the other one via these set of operations. There are six  classes in three-qubit pure states - a set of fully separable states, three biseparable classes and two SLOCC inequivalent  entangled classes, called the GHZ- and the W-class which are genuinely multipartite entangled. A new entanglement monotone in the context of monogamy of entanglement was introduced, called tangle \cite{dur00}, which can be defined for a tripartite state, $\rho_{ABC}$ as 
\begin{equation}
\mathcal{T} = \mathcal{C}^{2}_{A|BC} -  \mathcal{C}^{2}_{A|B} -  \mathcal{C}^{2}_{A|C}
\label{eq:tangle}
\end{equation}
where $\mathcal{C}_{A|B}$ and $\mathcal{C}_{A|C}$ are the concurrences \cite{hill97,wootters98} for the reduced density matrices, $\rho_{AB}$ and $\rho_{AC}$  respectively while $C_{A|BC}$ represents the concurrence of $\rho_{ABC}$ in the $A:BC$ bipartition. Note that for pure states, $\mathcal{C}^2_{A|BC}=4\det \rho_{A}$, where $\rho_A$ is the local density matrix of a pure state, $\ket{\psi_{ABC}}$. It can be easy to find that the states belonging to the GHZ-class have non vanishing tangle. An arbitrary state belonging to the three-qubit W-class state takes the form as
\(\ket{\psi}_W= \lambda_{0}\ket{000} + \lambda_{1}\ket{100} + \lambda_{2}\ket{001} + \lambda_{3}\ket{010}\)
for which tangle vanishes. Note, however, that the tangle also vanishes for separable and bi-separable classes of states. To distinguish between separable, bi-separable and the W-class states, we also require genuine multiparty entanglement measure, like generalized geometric measure which is nonvanishing only for the W-class states among the three classes. Therefore, the prescription for classifying the resulting three-qubit states, obtained both from biased as well as unbiased inflation processes can now be updated - if $\mathcal{G}^B_{\max}$ or $\mathcal{G}^{UB}_{\max}$ is nonvanishing after maximization, we compute tangle which can identify two classes.

To examine the role of ranks of unsharp measurement operators on the properties of the final state, let us first start the investigation for a class of initial state, namely $\ket{\psi_{in}}=\cos{\theta_s}\ket{00}+\sin \theta_s \ket{11}$ and the unsharp measurement described in Eqs. (\ref{eq:rank2povm})-(\ref{eq:rank4povm}). In this situation, we can have the following theorem.\\

\noindent\textbf{$\blacksquare$ Theorem I.}
\label{theorrem1}
\emph{In the biased as well as unbiased generation protocols of three-qubit states, for a fixed class of unsharp measurement originated from the Bell basis in Eqs. (\ref{eq:rank2povm})-(\ref{eq:rank4povm}), and for input states of the form $\ket{\psi_{in}}=\cos{\theta_s}\ket{00} + \sin{\theta_s}\ket{11}$, rank-$3$ measurement can produce both the GHZ- and the W-class states although  rank-$2$ and rank-$4$ unsharp measurements can lead to only the GHZ- and the W-class states respectively.}\\
\begin{proof}
Taking the initial two-qubit entangled state, $\ket{\psi_{in}}_{AB}=\cos{\theta_s}\ket{00} + \sin{\theta_s}\ket{11}$, and the auxiliary state, $\ket{\psi_{aux}}_C=\alpha\ket{0}+\beta\ket{1}=\cos{\theta_a/2}\ket{0}+e^{i\phi}\sin{\theta_a/2}\ket{1}$, we apply $\{M_k^{r=2}\}_{k=1}^4$ on the qubit $B$ and the auxiliary qubit $C$.  The three-qubit resulting state corresponding to the first outcome, \(M_1^2\), reads 
\begin{eqnarray}
\nonumber
&&\ket{\psi_{out,1}^2}_3 = \\ \nonumber
&&\frac{1}{2 \sqrt{N_2}}\big[\alpha \cos{\theta_s}\big\{(\sqrt{p}+\sqrt{1-p})\ket{000} + (\sqrt{p}-\sqrt{1-p})\ket{011}\big\}\\ \nonumber
&&+\beta \sin{\theta_s}\big\{(\sqrt{p}+\sqrt{1-p})\ket{111} + (\sqrt{p}-\sqrt{1-p})\ket{100}\big\}\big],\\
\end{eqnarray}
where $N_2 = \frac{1}{4}\{1+(|\alpha|^2-|\beta|^2)\cos{2\theta_s}\}$ is the normalization constant. Note that the superscript of the output state is for the rank of the measurement operator, while \(1\) in the subscript signifies the first outcome and  \(3\) represents the three-qubit output state.
Here $\mathcal{C}_{A|B} = \mathcal{C}_{A|C} = 0$ for this state and hence we find the tangle to be
\begin{eqnarray}
\mathcal{T}_{1}^2 = 4 \det \rho_A =4 \frac{(1-p)p \sin^2{\theta_a} \sin^2{2\theta_s}}{(1+\cos{\theta_a}\cos{2\theta_s})^2},
\label{eq:tangle1}
\end{eqnarray}
where the superscript of tangle signifies the rank of the measurement operator. 
If $\theta_s$ takes value $0$ or $\frac{\pi}{2}$,  the state becomes bi-separable leading to a vanishing GGM. For all other choices of \(\theta_s\), we can guarantee that $\mathcal{T} > 0$ when $\mathcal{G} \neq 0$, thereby creating the GHZ-class states. Such a conclusion can be arrived even without optimizing $\mathcal{G}$. Hence the results remain true both for biased and unbiased methods. 

When the output of the unsharp measurement is $M_1^{3}$, the output state takes the form as
\begin{eqnarray}
\nonumber
&&\ket{\psi_{out,1}^3}_3 = \\ \nonumber
&&\frac{1}{2\sqrt{N_3}}\bigg[\sqrt{\frac{p}{2}}\big\{(\alpha \cos{\theta_s}\ket{0}+\beta \sin{\theta_s}\ket{1})\ket{\phi^+} \\ \nonumber 
&&+(\alpha \cos{\theta_s}\ket{0}-\beta \sin{\theta_s}\ket{1})\ket{\phi^+}\big\}\\ 
&&+\sqrt{\frac{1-2p}{2}}(\beta \cos{\theta_s}\ket{0}+\alpha \sin{\theta_s}\ket{1})\ket{\psi^+}\bigg],
\label{eq:rank3state}
\end{eqnarray}
with normalization factor $N_3=\frac{1}{4}\{1+(4p-1)(|\alpha|^2-|\beta|^2)\cos{2\theta_s}\}$. Now, to calculate tangle, we find 
\begin{eqnarray}
\mathcal{C}_{A|BC}^2=\frac{8p\{5p-3+(3p-1)\cos{2\theta_a}\}\sin^2{2\theta_s}}{(2+2(4p-1)\cos{\theta_a}\cos{2\theta_s})^2},
\end{eqnarray}
and 
\begin{eqnarray}
\mathcal{C}_{AB}^2+\mathcal{C}_{AC}^2 = \frac{8p(1-2p)\sin^2{2\theta_s}}{(1+(4p-1)\cos{\theta_a}\cos{2\theta_s})^2}.
\end{eqnarray}
Therefore, the tangle of the output state after performing rank-$3$ measurement is given by
\begin{eqnarray}
\mathcal{T}_{1}^3=\frac{4p(3p-1)\sin^2{\theta_a}\sin^2{2\theta_s}}{(1+(4p-1)\cos{\theta_a}\cos{2\theta_s})^2}.
\end{eqnarray}
If $\theta_s = 0$ or $\pi/2$, $\mathcal{T}_{1}^3=0$, although in this case, $\mathcal{G}=0$ too, as can be seen from Eq. (\ref{eq:rank3state}), i.e., the output state becomes separable in the $A:BC$ bipartition. On the other hand, when $\theta_a=0$ or $\pi$, the tangle again vanishes although the Schmidt rank $(\vec{r_s})$ of the output state in all the partition is $\vec{r_s} = (r_{s(A)},r_{s(B)},r_{s(C)}) = (2,2,2)$. It means that the state remains genuine multiparty entangled with $\mathcal{G}\neq 0$. Except these parameter values, tangle can be non-zero. Hence, rank-$3$ unsharp measurements can create both the GHZ- and the W-class states.

Let us now move to the scenario of rank-$4$ unsharp measurements $\{M_k^{4}\}$ in Eq. (\ref{eq:rank4povm}). When $M_1^{4}$ clicks, the output state becomes
\begin{eqnarray}
\nonumber
&&\ket{\psi_{out,1}^4}_3 = \\ \nonumber
&&\frac{1}{2\sqrt{N_4}}\bigg[\sqrt{\frac{p}{2}}\big\{(\alpha \cos{\theta_s}\ket{0}+\beta \sin{\theta_s}\ket{1})\ket{\phi^+} \\ \nonumber 
&&+(\alpha \cos{\theta_s}\ket{0}-\beta \sin{\theta_s}\ket{1})\ket{\phi^+}\\ \nonumber
&&+(\beta \cos{\theta_s}\ket{0}+\alpha \sin{\theta_s}\ket{1})\ket{\psi^+}\big\}\\ 
&&+\sqrt{\frac{1-3p}{2}}(\beta \cos{\theta_s}\ket{0}-\alpha \sin{\theta_s}\ket{1})\ket{\psi^-}\bigg],
\label{eq:rank4state}
\end{eqnarray}
with $N_4=N_3$. The individual terms in the tangle are given by 
\begin{eqnarray}
\mathcal{C}_{A|BC}^2 &=& \frac{8p(1-2p)\sin^2{2\theta_s}}{(1+(4p-1)\cos{\theta_a}\cos{2\theta_s})^2},\\
\mathcal{C}_{A|B}^2 &=& \frac{4(1+2\sqrt{p(1-3p)}-2p)p\sin^2{2\theta_s}}{(1+(4p-1)\cos{\theta_a}\cos{2\theta_s})^2},\\
\mathcal{C}^2_{A|C} &=& \frac{4(1-2\sqrt{p(1-3p)}-2p)p\sin^2{2\theta_s}}{(1+(4p-1)\cos{\theta_a}\cos{2\theta_s})^2}.
\end{eqnarray}
For the entire range of parameters, we find $\mathcal{T}_{1}^4=0$ where $\mathcal{G} \neq 0$, thereby confirming the W-class states. In a similar fashion, for all the other outputs of rank-$2,3,4$ measurement operators, we can show that rank-$2$ and rank-$4$ measurements generate states which belong to the GHZ- and  W-class states respectively while rank-$3$ unsharp measurements can produce both of these two classes.
\end{proof}

\textbf{Remark 1. }
The results show that the set of unsharp measurements can  produce   three-qubit states having distinct features. It turns out that the classes generated by this method coincide with the known classification of three-qubit states \cite{dur00}.  

Moreover, the results demonstrate the power of unsharp measurements in the creation of different classes of genuine multipartite entangled states although the classifications obtained here are based on a specific class of unsharp measurement.

We will now examine whether the dependence of properties of the final state on the rank of the measurement operators reported in Theorem I is generic or not. To analyze it, we Haar uniformly generate two-qubit pure states of the form $\ket{\psi_{in}}=a_0\ket{00}+a_1\ket{01}+a_2\ket{10}+a_3\ket{11}$ where $a_i$s are complex numbers, and their real and imaginary parts are chosen randomly form Gaussian distribution with mean $0$ and unit standard deviation \cite{bengtsson_zyczkowski_2006}.
Following the protocol discussed in Sec. \ref{sec:protocol}, we can  calculate the three-party state according to the outcome $M_1^2$ as
\begin{eqnarray}
 \nonumber
 &&\ket{\psi_{out,1}^2}_3= \mathbb{I}\otimes\sqrt{M^2_{1}}\big(\ket{\psi_{in}}\otimes\ket{\psi_{aux}}\big)\\\nonumber
 &&=\alpha(a_0\ket{0} +a_2\ket{1})( \frac{\sqrt{p}+\sqrt{1-p}}{2}\ket{00}+ \frac{\sqrt{p}-\sqrt{1-p}}{2}\ket{11}) \\\nonumber
 &&+ \beta(a_1\ket{0} +a_3\ket{1})( \frac{\sqrt{p}+\sqrt{1-p}}{2}\ket{11}+ \frac{\sqrt{p}-\sqrt{1-p}}{2}\ket{00}),\\
 \label{eq:haar_m21}
\end{eqnarray}
where $\alpha$ and $\beta$ are parameters of the auxiliary state. Similarly, if $M^{2}_{2}$ clicks, we can get the output state as
\begin{eqnarray}
\nonumber
&&\ket{\psi_{out,2}^2}_3 \\\nonumber
&&= \frac{\sqrt{p}}{2}[(\alpha a_0- \beta a_1)(\ket{000}-  \ket{011})+(\alpha a_2 - \beta a_3 )(\ket{100} \\\nonumber &&-\ket{111})] + \frac{\sqrt{1-p}}{2}[(\beta a_0 + \alpha a_1)(\ket{010}+\ket{001})+ \\ &&(\alpha a_3 +\beta a_2)(\ket{101}+\ket{110})]. 
\label{eq:haar_m22}
\end{eqnarray}
Since $M^{2}_{3}$ and $M^{2}_{4}$ are local unitarily connected to $M^{2}_{1}$ and $M^{2}_{2}$ respectively, the final states remain the same as in Eqs. (\ref{eq:haar_m21})-(\ref{eq:haar_m22}) upto local unitary transformation. 
By considering unsharp measurement with rank-$3$ and -$4$, the resulting state can also be obtained for arbitrary measurement outcomes.

\begin{figure*}[ht]
\includegraphics[width=\textwidth]{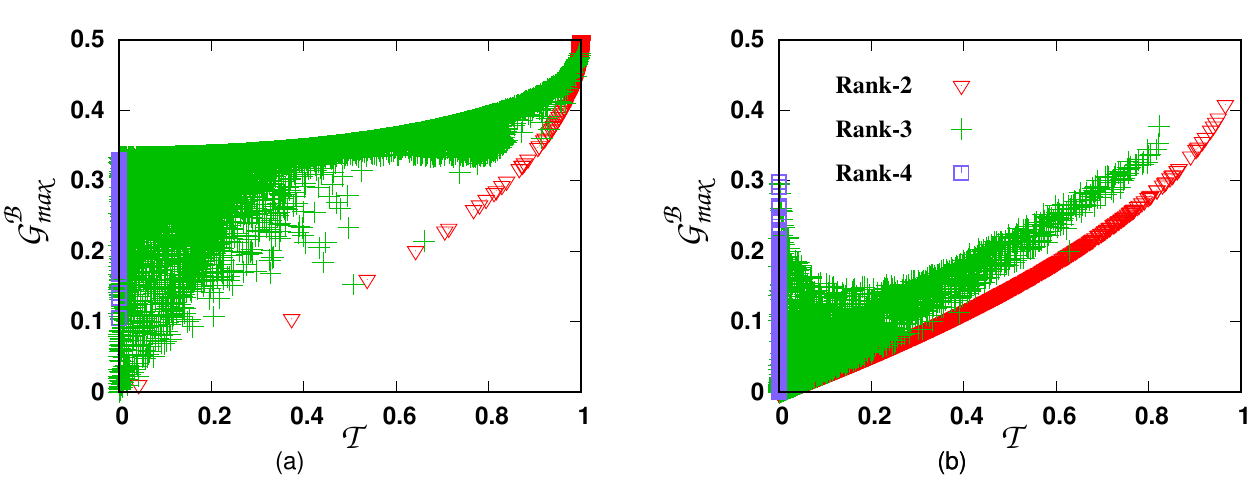}
\caption{(Color online.) 
{\bf Optimized GGM, \(\mathcal{G}_{\max}^B\), (ordinate) vs. tangle, \(\mathcal{T}\), (abscissa) in the biased inflation scheme.} 
(a) After optimizing the outcome $M_1^r$,  we plot the maximized GGM, $\mathcal{G}^B_{\max}$ against tangle, $\mathcal{T}$ of the generated three-qubit states for  unsharp measurement operators of rank-$2$, rank-$3$ and rank-$4$. We Haar uniformly generate \(5 \times 10^4\) two-qubit states and we choose an arbitrary auxiliary qubit  for which  state parameters are chosen in such a way that  GGM is maximized. 
(b) The similar plot when 
$M_2^r$ clicks where the previous optimized values for \(p\) and auxiliary state parameters are used. The resulting states after obtaining other two outcomes have qualitatively similar feature like \(M_2^r\).  Both the axes are dimensionless.}
\label{fig:ggm_vs_tangle_3} 
\end{figure*}

The scattered values of $\mathcal{G}_{\max}^B$ and tangle for the output states corresponding to the outcomes, \(M_1^r\) and \(M_2^r\), are depicted in Figs. \ref{fig:ggm_vs_tangle_3}(a) and (b). We observe that Theorem I holds even for generic states.
In particular, 
the three-qubit states obtained after rank-$4$ unsharp measurement lie in the line of $\mathcal{T}=0$ while rank-$3$ measurements produce states with both $\mathcal{T}=0$ and $\mathcal{T}\neq0$ with nonvanishing \(\mathcal{G}\). Notice that the features remain qualitatively similar for other outcomes of the measurement as shown in Fig. \ref{fig:ggm_vs_tangle_3}(b). 

\begin{figure}
\includegraphics[width=1.0\linewidth]{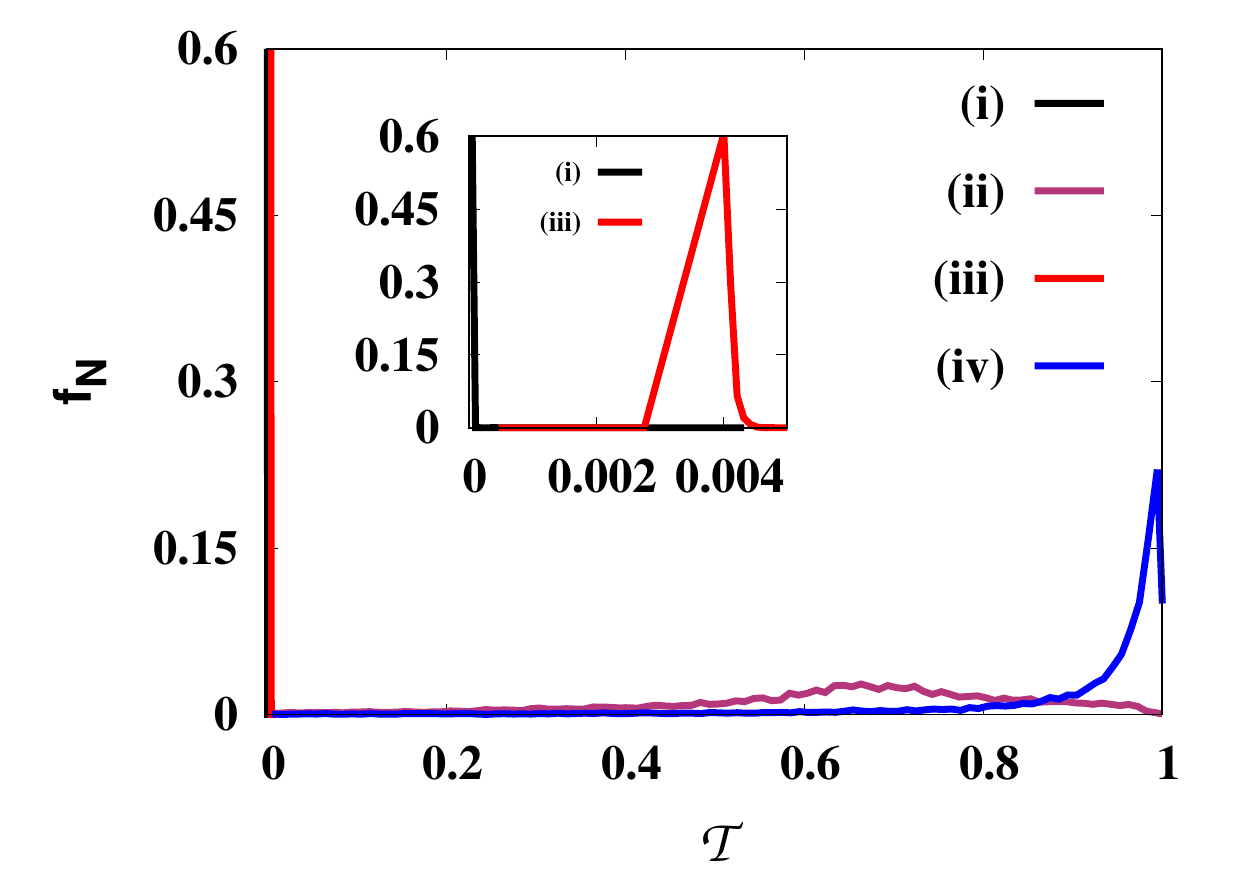}
\caption{(Color online.) Frequency distribution, \(f_N\) (vertical axis) of $\mathcal{T}$ 
(horizontal axis).  We Haar uniformly generate  \(10^4\) two-qubit states. Here (i) $\mathcal{T}_{\min}$ with rank-$3$, (ii) $\mathcal{T}_{\max}$ with rank-$3$, (iii) $\mathcal{T}_{\min}$ with rank-$2$, and  (iv) $\mathcal{T}_{\max}$ with rank-$2$ unsharp measurements are calculated for the three-qubit states which are produced via the inflation process. 
Both $\mathcal{T}_{\max}$ and $\mathcal{T}_{\min}$ vanish  for all the random states generated when rank-$4$ unsharp measurements are performed.
The optimizations are carried out as discussed in Eq. (\ref{tangle_max_min}).
(Inset) The cases (i) and (iii) are depicted. 
Both the axes are dimensionless. }
\label{fig:tmax_tmin} 
\end{figure}

To ensure Theorem I to be valid for Haar unifromly generated states numerically, we compute both maximum and minimum values of the tangle for a given state where the maximization is performed over auxiliary state parameters and control parameters of the unsharp measurements in Eqs. (\ref{eq:rank2povm})-(\ref{eq:rank4povm}). Precisely, for a given state and a fixed outcome $M_k^r$ with a constraint $\mathcal{G} \neq 0$, we find 
\begin{eqnarray}
\nonumber
\mathcal{T}_{\max} = \underset{\{p,\theta_a,\phi_a,\mathcal{G}\neq 0\}}{\mathrm{max}} \mathcal{T}\{ \sqrt{M_{1}}\ket{\psi_{in}}\otimes\ket{\psi_{aux}}\}, \\
\mathcal{T}_{\min} = \underset{\{p,\theta_a,\phi_a,\mathcal{G}\neq 0\}}{\mathrm{min}} \mathcal{T}\{ \sqrt{M_{1}}\ket{\psi_{in}}\otimes\ket{\psi_{aux}}\}.
\label{tangle_max_min}
\end{eqnarray}
We can see the frequency distribution of $\mathcal{T}_{\max}$ and $\mathcal{T}_{\min}$  of final three-qubit states inflated from Haar uniformly chosen  two-qubit states in Fig.~\ref{fig:tmax_tmin} for different rank of measurement operators. The observations from Fig. \ref{fig:tmax_tmin} is in good agreement with Theorem I. 
\begin{enumerate}
    \item First of all, we find that $\mathcal{T}_{\min}$ with rank-$3$ unsharp measurements can vanish and at the same time, $\mathcal{T}_{\max}$ can be nonvanishing when GGM is nonvanishing.
    
    \item The different picture emerges for rank-$2$ and rank-$4$ measurements. In case of rank-$2$ measurements, $\mathcal{T}_{\min}$ as well as $\mathcal{T}_{\max}$ never vanish, thereby ensuring the generation of the GHZ-class states. On the other hand, $\mathcal{T}_{\max}$ as well as $\mathcal{T}_{\min}$ both vanish, when rank-$4$ measurements are applied, thereby establishing the creation of the W-class states. It is true for both biased and unbiased scenarios. Therefore, we can safely arrive at the following Proposition. 
\end{enumerate}
\noindent\textbf{$\blacksquare$ Proposition I.}
\emph{In the  entanglement inflation procedure, we identify a class of unsharp two-qubit  measurement operators of different ranks which when acted on Haar uniformly generated input states and on an auxiliary qubit, can create both the GHZ- and W-class states when the rank of the measurement operator is three, although  only the GHZ- and the W-class states are produced for rank-$2$ and rank-$4$ measurements respectively.}
\label{proposition1}\\

Let us now analyze the optimized GGM, $\langle \mathcal{G} \rangle$ created by unsharp measurements of rank-$2$, -$3$ and -$4$ in the biased process. For brevity, we use $\langle \mathcal{G} \rangle$  instead of $\langle \mathcal{G}^B_{\max} \rangle$.  When we optimize GGM for a single outcome of the measurements, the GGM can achieve much higher value compared to that obtained for other outcomes (see Table. \ref{table:3qubit}).
\begin{widetext}
\begin{table*}[ht]
  \begin{tabular}{|c|c|c|c|c|}
\hline
& $M_{1}^r$ & $M_{2}^r$ & $M_{3}^r$ & $M_{4}^r$  \\
\hline
          \begin{tabular}{c}
            Rank of measurement\\
           \hline
          2 \\
          \hline
          3 \\
         \hline
          4 \\
          \end{tabular}
          &
          \begin{tabular}{c|c}
             $\langle\mathcal{G}\rangle$ & $\sigma_{\mathcal{G}}$\\ 
          \hline 
          $0.4961$    & $0.0111$   \\
          \hline
         $0.3353$  &$0.0547$    \\
            \hline
         $0.2617$  &$0.0319$    \\   
         \end{tabular}         
          &
          \begin{tabular}{c|c}
             $\langle\mathcal{G}\rangle$ & $\sigma_{\mathcal{G}}$\\ 
          \hline 
          $0.0369$    & $0.0502$   \\
          \hline
         $0.0546$  &$0.0655$    \\
            \hline
         $0.0290$  &$0.0422$    \\   
         \end{tabular}          
         &
          \begin{tabular}{c|c}
             $\langle\mathcal{G}\rangle$ & $\sigma_{\mathcal{G}}$\\ 
          \hline 
          $0.0708$    & $0.0904$   \\
          \hline
         $0.0649$  &$0.0876$    \\
            \hline
         $0.0289$  &$0.0424$    \\   
         \end{tabular} 
          &
          \begin{tabular}{c|c}
             $\langle\mathcal{G}\rangle$ & $\sigma_{\mathcal{G}}$\\ 
          \hline 
          $0.0370$    & $0.0503$   \\
          \hline
         $0.0518$  &$0.0610$    \\
            \hline
         $0.0289$  &$0.0422$    \\   
         \end{tabular} \\
\hline                      
\end{tabular}\\
\caption{ Mean and standard deviation of GGM   of the output three-qubit states  corresponding to different  outcomes of measurements with different ranks in the biased inflation protocol. Data generated for the analysis is \(5 \times 10^4\).}
\label{table:3qubit}
      
  \end{table*}
\end{widetext}
Moreover, we observe that $\langle \mathcal{G} \rangle$ decreases with increasing rank of the measurement while standard deviation, $\sigma_{\mathcal{G}}$ increases. Hence, it can be concluded that higher rank measurements are not suitable to create high genuine multipartite entangled states while low rank states are capable to produce highly genuine multipartite entangled states. It is interesting to study whether such observation remains true for higher dimensional joint measurements or states with higher number of parties.


\section{Generated Multiqubit state characterization via persistency, tangle and GGM}

\label{sec:multiqubit}

Let us move our attention to the generation of GME states containing more than three parties. Unlike three-qubit pure states, there exist infinite number of SLOCC inequivalent classes for multiqubit states. Hence, classification of states in this case is not easy even for pure states. One possibility is to characterize the multipartite states according to their usefulness in quantum information processing tasks. In this work, we present the classification of multiqubit states with the help of persistency, GGM and tangle.

\subsection{Persistent state generation}

The persistency quantifies how hard it is to destroy entanglement of a given state under local measurements and $0 \leq P_e \leq N-1$ for $N$-qubit states. It was shown that the persistency of cluster state, resource for one-way computer \cite{Rausendorf2001}, is $\lfloor \frac{N}{2} \rfloor$ \cite{persistency01}. Before presenting the results, let us define it.

\textit{Persistency of entanglement}. Let $\ket{\Psi_N}$ be a $N$-party pure entangled state. Persistency, $P_e$ is defined as the minimum number of single-qubit local measurements required to completely disentangle the state into a fully seprable state for all measurement outcomes. Higher values of persistency means that it is harder to destroy entanglement of the particular state by local measurements.

Suppose, we initially take $\ket{\psi_{in}}$ as a general $(N-1)$-qubit state and an auxiliary state. We now want to check whether we can build $N$-qubit GME states having different persistency by varying rank of the unsharp measurement operators.\\

\noindent\textbf{$\blacksquare$ Proposition II.} \emph{For arbitrary $(N-1)$-qubit initial resource state and a single-qubit state as an auxiliary  system, a class of unsharp measurement of rank-$2$ can generate $N$-party GME states with persistency, $P_e=N-2$, while both  rank-$3$, and -$4$ measurements lead to GME states with $P_e=N-1$.}
\begin{proof}
We start with an arbitrary three-qubit state, $\ket{\psi_{in}} = a_0\ket{000} + a_1\ket{001} + a_2\ket{010} + a_3\ket{011} + a_4\ket{100} + a_5\ket{101} + a_6\ket{110} + a_7\ket{111}$ and an auxiliary state $\ket{\psi_{aux}}=\alpha\ket{0}+\beta\ket{1}$ where $a_i$s $(i=0,1,\ldots,7)$, $\alpha$ and $\beta$ are complex numbers satisfying normalization condition. The initial state is $\ket{\psi_{in}}_{123}\otimes\ket{\psi_{aux}}_4$ and the rank-$2$ measurement is performed on the third and the fourth qubit. Let us first discuss the scenario for rank-$2$ unsharp measurements. If $M_1^2=p\ket{\phi^+}\bra{\phi^+} + (1-p)\ket{\phi^-}\bra{\phi^-}$ clicks, the post-measured state (upto some normalization factor) can be rewritten as
\begin{eqnarray}
\nonumber &\ket{\Psi_{out,1}^2}_{4}& \to \\ \nonumber
&\ket{\chi}_2 \ket{0}_{3}&\bigg[\cos{\frac{\theta_2}{2}}\ket{Y^+}_1 + e^{-i\phi_2}\sin{\frac{\theta_2}{2}}\ket{Z^+}_1\bigg]\ket{0}_4\\ \nonumber
&+ \ket{\chi}_2 \ket{1}_{3}&\bigg[\cos{\frac{\theta_2}{2}}\ket{Y^-}_1 + e^{-i\phi_2}\sin{\frac{\theta_2}{2}}\ket{Z^-}_1\bigg]\ket{1}_4\\ \nonumber
&+ \ket{\chi^\perp}_2 \ket{0}_{3}&\bigg[\cos{\frac{\theta_2}{2}}\ket{Y^+}_1 - e^{-i\phi_2}\sin{\frac{\theta_2}{2}}\ket{Z^+}_1\bigg]\ket{0}_4\\ \nonumber
&+ \ket{\chi^\perp}_2 \ket{1}_{3}&\bigg[\cos{\frac{\theta_2}{2}}\ket{Y^-}_1 - e^{-i\phi_2}\sin{\frac{\theta_2}{2}}\ket{Z^-}_1\bigg]\ket{1}_4,\\
\end{eqnarray}
where 
\begin{eqnarray}
\nonumber \ket{\chi}_k &=& \cos{\frac{\theta_k}{2}}\ket{0}_k + e^{i\phi_k}\sin{\frac{\theta_k}{2}}\ket{1}_k,\\
\ket{\chi ^ \perp}_k &=& \sin{\frac{\theta_k}{2}}\ket{0}_k - e^{i\phi_k}\cos{\frac{\theta_k}{2}}\ket{1}_k,
\end{eqnarray}
and
\begin{eqnarray}
\nonumber \ket{Y^\pm}&=&\sqrt{p}\big(\alpha\ket{A}+\beta\ket{B}\big) \pm \sqrt{1-p}\big(\alpha\ket{A}-\beta\ket{B}\big),\\
\nonumber \ket{Z^+}&=&\sqrt{p}\big(\alpha\ket{C}+\beta\ket{D}\big) \pm \sqrt{1-p}\big(\alpha\ket{C}-\beta\ket{D}\big),\\
\end{eqnarray}
with
\begin{eqnarray}
\nonumber \ket{A} &=& a_0\ket{0}+a_4\ket{1}\hspace{0.1cm};\hspace{0.1cm} \ket{B}=a_1\ket{0}+a_5\ket{1},\\
\ket{C} &=& a_2\ket{0}+a_6\ket{1}\hspace{0.1cm};\hspace{0.1cm} \ket{D}=a_3\ket{0}+a_7\ket{1}.
\end{eqnarray}
For computing persistency from the above form of the state, it is clear that if one measures the second qubit in the basis $\{\ket{\chi}_2, \ket{\chi^\perp }_2\}$ and  the third qubit in the computational basis, the state becomes fully unentangled for any possible outcome. In fact, measuring in any arbitrary direction on one of the qubits where unsharp measurement is not performed and in $\sigma_z$ basis on another qubit on which unsharp measurement is applied, the state becomes fully separable. Similarly, taking initial state as general four-qubit state, we can create five-qubit genuine multipartite entangled states. It can be shown that measuring two qubits from those qubits on which measurement in inflation is not performed, and measuring a qubit between the two on which unsharp measurement is applied, the state becomes fully separable for rank-$2$ unsharp measurement operators. Moreover, we check numerically that no less than two and three single-qubit local operations are required to make the GME states fully disentangled for Haar unifromly generated three- and four-qubit initial states respectively. Therefore, four- and five-qubit states generated in the protocol by rank-$2$  unsharp measurements have persistency $2$ and $3$ respectively. 

Looking at the output states for four- and five-qubit states, one can obtain a recursion relation of the $N$-qubit resulting state via rank-$2$ measurement, given by 
\begin{eqnarray}
\nonumber
\ket{\Psi_N^1} = &\sum&_{\mathcal{K}=0}^{2^{N-4}-1}\ket{\mathcal{K}}_{2,\ldots,N-3} \big\{ \ket{00}_{N-2,N-1}\ket{Y_{\mathcal{K}}^+}_1 \ket{0}_N \\ \nonumber &+& \ket{01}\ket{Y_{\mathcal{K}}^-}\ket{1} + \ket{10}\ket{Z_{\mathcal{K}}^+}\ket{0}+\ket{11}\ket{Z_{\mathcal{K}}^-}\ket{1} \big\},\\
\end{eqnarray}
where  $\mathcal{K}$ is the decimal value of the computational basis formed by $2, \ldots, N-3$ qubits. And, we have 
\begin{eqnarray}
\nonumber
\ket{Y_{\mathcal{K}}^\pm} &=& \sqrt{p}\big(\alpha\ket{A_{4\mathcal{K}}}+\beta\ket{A_{4{\mathcal{K}}+1}}\big)\\ \nonumber
&\pm& \sqrt{1-p}\big(\alpha\ket{A_{4\mathcal{K}}}-\beta\ket{A_{4{\mathcal{K}}+1}}\big),\\ \nonumber
\ket{Z_\mathcal{K}^\pm} &=& \sqrt{p}\big(\alpha\ket{A_{4\mathcal{K}+2}}+\beta\ket{A_{4{\mathcal{K}}+3}}\big)\\ 
&\pm& \sqrt{1-p}\big(\alpha\ket{A_{4\mathcal{K}+2}}-\beta\ket{A_{4{\mathcal{K}}+3}}\big), 
\end{eqnarray}
where
\begin{eqnarray}
\nonumber
A_i = a_i\ket{0} + a_{\frac{2^{N-1}}{2}+i}\ket{1},\hspace{0.2cm}i=0, 1,\ldots, (2^{N-1}-1).\\
\end{eqnarray}
Our strategy is to measure $2, 3, \ldots, N-2 $ and $(N-1)$th qubit in computational basis to destroy the entanglement of the state. Although, it can be checked that one can also measure the first $2, 3, \ldots, N-2 $ qubits in arbitrary direction. The basis of the output state created for the outcome $M_3^2$ is local unitarily connected to the basis of the output for $M_1^2$ due the local equivalence of these two measurement elements in rank-$2$ measurements. Therefore, the same strategy can be applied here to disentangle the state fully. Thus, depending on the numerical analysis for four- and five-qubit generated states and from the recursion relation, we prove that for $N$-qubit GME state, $P_e=N-2$ when rank-$2$ unsharp measurement operators  are used to  obtain $N$-qubit entangled states.
Again, $M_2^2$ and $M_4^2$ outcomes are local unitarily connected. In this case, if one measures the qubits of position $2,3, \ldots, N-2$ in arbitrary direction or simply in the computational basis and $N-1$-th qubit in $\theta=\pi/2, \phi = \pi/2$ direction, i.e., in the $\sigma_y$ basis, the state gets disentangled, which gives persistency $P_e = N-2$.

For rank-$3$ and -$4$ unsharp measurements, one can  find that all $(N-1)$ local measurements can lead to fully separable states from GME state produced via the inflation process. Numerical simulations indicate that four- and five-qubit generated states obtained after rank-$3$ and -$4$ measurements can not be disentangled via local operations performed on two and three qubits respectively, thereby implying $P_e =N-1$.
\end{proof}

Since the output states obtained after  rank-$3$ and rank-$4$ unsharp measurements cannot be distinguished from the persistency  and hence it is interesting to see whether they can be distinguished with the help of GGM and tangle or not. In the succeeding subsection, we will address this issue. 

\textbf{Remark 1.} From the numerical simulations, we observe that instead of tripartite state as resource, starting with bipartite entangled states  and two auxiliary qubits, if one generates four-qubit state by applying unsharp measurements twice, the persistency of the produced state still remains two which is in agreement with  Proposition II. 

\textbf{Remark 2.} 
In Proposition II, we have reported that rank-$2$ unsharp measurement can lead to four-qubit states with persistency two. Moreover, we know that  the persistency of the cluster state, given by $\ket{\psi_c}=\frac{1}{2}(\ket{0000}+\ket{0011}+\ket{1100}-\ket{1111})$ is also two. Hence, it is tempting to ask whether the generated state is close to the cluster state or not for some sets of unsharp measurements. Specifically, we take a generalized GHZ state, given by $\ket{gGHZ}=\cos{\theta_s}\ket{000}+\sin{\theta_s}\ket{111}$ and a single-qubit auxiliary system as before, thereby obtaining the initial state as $\ket{gGHZ}_{123}\otimes\ket{\psi_{aux}}_{4}$.  We apply the rank-$2$ unsharp measurement on the third and fourth qubit of the initial state 
given by
\begin{eqnarray}
\nonumber
\big\{&&M_1^2 = p \mathbf{P}\big[\alpha\ket{00}+\beta\ket{11}\big] + (1-p) \mathbf{P}\big[-\beta^*\ket{00}+\alpha\ket{11}\big],\\
&&\nonumber M_2^2 = p \mathbf{P}\big[-\beta^*\ket{00}+\alpha\ket{11}\big] + (1-p) \mathbf{P}\big[\alpha\ket{01}+\beta\ket{10}\big],\\
&&\nonumber M_3^2 = p \mathbf{P}\big[\alpha\ket{01}+\beta\ket{10}\big] + (1-p) \mathbf{P}\big[-\beta^*\ket{01}+\alpha\ket{10}\big],\\
&&\nonumber M_4^2 = p
\big[\mathbf{P}\big[-\beta^*\ket{01}+\alpha\ket{10}\big] + (1-p) \mathbf{P}\big[\alpha\ket{00}+\beta\ket{11}\big]\big\},\\
\end{eqnarray}
where $0 \leq p \leq 1$ and $\alpha = \cos{\zeta}$,  $\beta=e^{-i\xi}\sin{\zeta}$ satisfying normalization constraints. Here \(\mathbf{P}\big[\alpha\ket{00}+\beta\ket{11}\big] = (\alpha\ket{00}+\beta\ket{11}) (\alpha\bra{00}+\beta^*\bra{11}) \).
We can write the first element as
\begin{eqnarray}
\nonumber\sqrt{M_1^2} &=& C\ket{00}\bra{11} + D\ket{00}\bra{11} \\
&+& {D}^* \ket{11}\bra{00} + E\ket{11}\bra{11},\\ \nonumber
\text{where}\\ \nonumber
C &=& \frac{1}{2}\{\sqrt{1-p}(1-\cos{\zeta})+\sqrt{p}(1+\cos{\zeta})\},\\ \nonumber
D &=& \frac{1}{2}e^{-i\xi}\sin{\zeta}(\sqrt{p} - \sqrt{1-p}),\\ \nonumber
E &=& \frac{1}{2}\{\sqrt{p}(1-\cos{\zeta})+\sqrt{1-p}(1+\cos{\zeta})\}. \\
\end{eqnarray}
Following the inflation protocol to create four-qubit states, when $M_1^2$ clicks, the generated state takes the form
\begin{eqnarray}
\nonumber
    \ket{\psi_{out,1}^2}_4 \to \frac{1}{2}\bigg[&C& \alpha a_0\ket{0000} + D \beta a_7\ket{1100} \\
    &&+ {D}^* \alpha a_0 \ket{0011} + E \beta a_7 \ket{1111}\bigg],
\end{eqnarray}
where $a_0 = \cos{\theta_s}$ and $a_7 = \sin{\theta_s}$. From the above expression, it is clear that the basis of a cluster state matches with $\ket{\psi_{out,1}^2}_4$. We now maximize the fidelity between the produced state and the cluster state, i.e., $\mathcal{F}_m = \underset{p,\alpha,\beta,\theta_a, \phi_a}{\max}\big|_4\big\langle{\psi_{out,1}^2}\big|\psi_c\big\rangle\big|^2$ where the maximization is performed over the parameters involved in measurements and the auxiliary system for a fixed initial state parameter, $\theta_s$. We find that $\mathcal{F}_m$ can go around $0.73$ for the first outcome. Also, if $M_2^2$ clicks, then after applying $\sigma_x$ operation to the fourth qubit, we can recover the basis of the cluster state with different coefficients whose $\mathcal{F}_m < 0.73$. For other outcomes, fidelities can also be calculated after optimizing over local measurements. 

\subsection{Discrimination of rank-$3$ and rank-$4$ measurements via GGM and tangle}
\label{sec:tangle-ggm}

\begin{figure*}[ht]
\includegraphics[width=\textwidth]{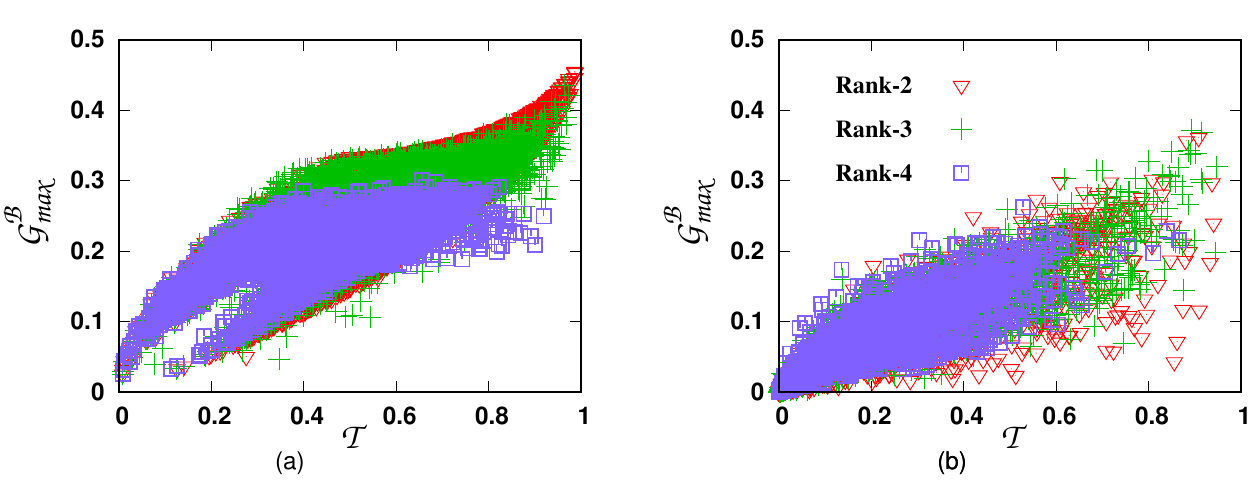}
\caption{(Color online.) \textbf{GGM (\(y\)-axis) against tangle (\(x\)-axis) for four-qubit  states created via the biased inflation protocol.} All other specifications are same as in Fig. \ref{fig:ggm_vs_tangle_3}.
Both the axes are dimensionless.}
\label{fig:ggm_vs_tangle_4} 
\end{figure*}

Let us now Haar uniformly generate three-qubit pure states and perform rank-$2$, -$3$ and -$4$ unsharp measurements given in Eq. (\ref{eq:rpovm}) on one of the qubits and an arbitrary auxiliary qubit. Hence we maximize the GGM of the output state obtained when the outcome is $M_1^r$. Clearly, $M_1^3$ and $M_1^2$ produce states with high GGM and high tangle as shown in Fig. \ref{fig:ggm_vs_tangle_4}.
Precisely, in the $(\mathcal{G}, \mathcal{T})$-plane, they can  not be distinguished easily although for the measurement outcome  $M_1^4$, states possesses less GGM and tangle on average (see Table. \ref{table:4qubit}) which is also in good agreement with the results for three-qubit output states in Sec. \ref{sec:3qubit}. Qualitatively similar results can be obtained when $M_i^r$ $(i \neq 1)$ clicks in the inflation protocol as depicted in Fig. \ref{fig:ggm_vs_tangle_4}(b) (similar to Fig. \ref{fig:ggm_vs_tangle_3}(b)).

\begin{widetext}
\begin{table*}[ht]
  \begin{tabular}{|c|c|c|c|c|}
\hline
& $M_{1}^r$ & $M_{2}^r$ & $M_{3}^r$ & $M_{4}^r$  \\
\hline
          \begin{tabular}{c}
           Rank of measurements \\
           \hline
          2 \\
          \hline
          3 \\
         \hline
          4 \\
          \end{tabular}
          &
          \begin{tabular}{c|c}
             $\langle\mathcal{G}\rangle$ & $\sigma_{\mathcal{G}}$\\ 
          \hline 
          $0.2765$    & $0.0724$   \\
          \hline
         $0.2618$  &$0.0637$    \\
            \hline
         $0.2096$  &$0.0369$    \\   
         \end{tabular}         
          &
          \begin{tabular}{c|c}
             $\langle\mathcal{G}\rangle$ & $\sigma_{\mathcal{G}}$\\ 
          \hline 
          $0.0593$    & $0.0542$   \\
          \hline
         $0.0827$  &$0.0551$    \\
            \hline
         $0.0573$  &$0.0451$    \\   
         \end{tabular}          
         &
          \begin{tabular}{c|c}
             $\langle\mathcal{G}\rangle$ & $\sigma_{\mathcal{G}}$\\ 
          \hline 
          $0.1012$    & $0.0783$   \\
          \hline
         $0.0959$  &$0.0738$    \\
            \hline
         $0.0571$  &$0.0455$    \\   
         \end{tabular} 
          &
          \begin{tabular}{c|c}
             $\langle\mathcal{G}\rangle$ & $\sigma_{\mathcal{G}}$\\ 
          \hline 
          $0.0594$    & $0.0543$   \\
          \hline
         $0.0822$  &$0.0595$    \\
            \hline
         $0.0576$  &$0.0466$    \\   
         \end{tabular} \\
\hline                      
\end{tabular}\\
\caption{ Mean and standard deviation of GGM of the resulting four-qubit states  corresponding to different unsharp measurement operators with different ranks. Data generated for the analysis is \(5 \times 10^4\) in the biased inflation scheme.}
\label{table:4qubit}
      
  \end{table*}
\end{widetext}

 \section{Unbiased inflation state production}
\label{subsec:unbiased state gen}

From the definition of the biased inflation protocol, it is clear that the protocol has a preference towards the post-selection $M_{k}^r$ which can also be seen from Fig.~\ref{fig:ggm_vs_tangle_3}. The mean GGM accessible through the output state when $M_{k}^r$ clicks is higher than the other post-selected states for different outcomes of the measurement.  

\begin{figure*}[ht]
\includegraphics[width=\textwidth]{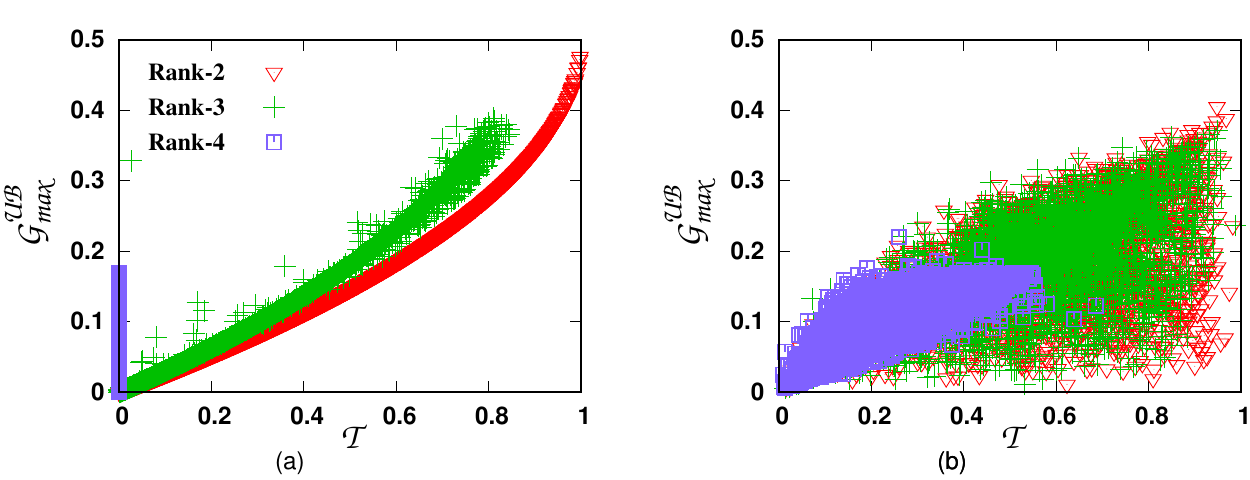}
\caption{(Color online.) Characteristics of optimized  GGM, $\mathcal{G}_{\max}^{UB}$ (vertical axis) against tangle $\mathcal{T}$ (horizontal axis)  for  unbiased inflation protocol. (a) The features are for  three-qubit output states produced from random two-qubit states, (b) the same is plotted for four-qubit final states from initial three-qubit states. Both the axes are dimensionless.}
\label{fig:pigi3-4qubit} 
\end{figure*}

Let us demonstrate how statistical characteristics of GGM for the output states after averaging over the various post-selections changes compared to that of the biased ones.
To compare between biased and unbiased entanglement inflation protocols, we again generate randomly $10^{4}$ two- and three-qubit initial states. We then compute GGM and tangle of the created states for each outcome and  optimize average GGM over measurement parameter, \(p\) and auxiliary state parameters as defined in Eq. (\ref{protocol2}).  
We again find that the three-qubit states produced  belong only to the GHZ- and the W-class  when the rank of weak measurement operators are two and four respectively while both the GHZ- and  W-class states are created when the weak measurement is of rank-$3$ (see Fig.~\ref{fig:pigi3-4qubit}). Both the average GGM and its standard deviation for the frequency distribution of the resulting states decreases with the increase in the rank of the unsharp measurements (as shown in Table.~\ref{tab:unbiased}). The similar feature of $\langle \mathcal{G}\rangle$, and  $\sigma_{\mathcal{G}}$ is also found for four-qubit  states which are created by generating three-qubit random states.  It is evident from the analysis that the qualitative features of the produced states in the unbiased process are same as in the biased ones. The only difference is that  GGM is typically lower for the unbiased protocol than that of the biased case with a mesurement outcome which is used for optimization. 


\textbf{Observation.} Interestingly, in the unbiased protocol, the optimal auxiliary state for the rank-$2$ and -$3$ unsharp measurements can be found.   Specifically, we find that $\theta_a \approx \pi/2$, $\phi_a \approx 0.0$. Moreover,  the optimal tuning parameter for rank-$2$  measurement is  found to be $ \approx 0.5$ while  for rank-$3$, $p\approx0.48$ for all the Haar uniformly generated states. Moreover, we notice that all \(\mathcal{G}_k\)s  in Eq. (\ref{protocol2}) are equal  with equal probabilities \(q_k\) in these scenarios. 
Such universal optimal value for \(p\) is not observed  for unsharp measurements having rank-$4$ although \(\mathcal{G}_k\)s and \(q_k\)s are  almost equal. 


\begin{table}
  \begin{tabular}{|c|c|c|c|c|}
 \hline 
& $3$-qubit & $4$-qubit \\
\hline
          \begin{tabular}{c}
           Rank of the measurement \\
           \hline
          2 \\
          \hline
          3 \\
         \hline
          4 \\
          \end{tabular}
          &
          \begin{tabular}{c|c}
             $\langle\mathcal{G}\rangle$ & $\sigma_{\mathcal{G}}$\\ 
          \hline 
          $0.1621$    & $0.0691$   \\
          \hline
         $0.1614$  &$0.0681$    \\
            \hline
         $0.1066$  &$0.0370$    \\   
         \end{tabular}         
          &
          \begin{tabular}{c|c}
             $\langle\mathcal{G}\rangle$ & $\sigma_{\mathcal{G}}$\\ 
          \hline 
          $0.1255$    & $0.0968$   \\
          \hline
         $0.1233$  &$0.0923$    \\
            \hline
         $0.0748$  &$0.0458$    \\   
         \end{tabular} \\
\hline                      
\end{tabular}\\

\caption{ Mean and standard deviation of GGM for the resulting three- and four-qubit states  corresponding to a single outcome of the unsharp measurements with different ranks where the optimizatioin is performed accoroding to the unbiased inflation protocol in Eq. (\ref{protocol2}). The initial \(5 \times 10^4\) two- and three-qubit states are generated Haar uniformly. Point to note that the statistical behavior of  GGM for the output states obtained in this unbiased protocol is independent of different post-selections. }
 \label{tab:unbiased}     
  \end{table}

\section{Optimal choice of resource}
\label{subsec:optimal_resource_distribution}

The entanglement inflation protocol is the one by which genuine multiparty entangled states can be created from an entangled state with a less number of parties and with the help of the auxiliary system. In this case, the optimal resource state which can be used as an initial state is crucial. The question of finding optimal initial state  will be addressed here.\\
Suppose there is a genuinely multiparty entangled state of, say, $N_{1}$-party and $N_{2}$ number of single-qubit auxiliary systems are available. With the help of global weak measurement of different ranks on one party of the entangled state and the auxiliary qubit, one can generate a genuinely  multiparty entangled state of $(N_{1} + 1) $ parties. Repeating the process $(N_{2}-1)$ times, we can create a multiparty entangled state of $(N_{1}+ N_{2})$ parties. Hence finding the optimal system-size $N_{1}$ which leads to a high genuine multiparty entangled state can be an interesting issue to address.
Specifically, to create a $N$-party  entangled state, having  $\mathcal{G}$, the system size of the initial entangled state of $N_{1}$-party $(N_{1}\leq N-1)$ is important to determine. Let us illustrate the situation,  when $N$ = 5 which  have four possibilities:
\begin{itemize}
    \item \textbf{Scenario 1.}  Let us first prepare a two-qubit entangled state and three single-qubit auxiliary states.  We apply three weak measurements to produce five-qubit entangled state. The configuration can be denoted as $(2+1+1+1)$.
    \item \textbf{Scenario 2 and 3.} A three-qubit genuinely multipartite entangled state chosen either from the GHZ- or the W-class and two single-qubit auxiliary states with two weak measurements can produce five-qubit entangled state referring to $(3+1+1)$ scenario.
    
    \item \textbf{Scenario 4.} Finally, a four-qubit genuinely entangled state can create a five-qubit entangled state with the help of  single-qubit auxiliary state and a two-qubit weak measurement which we call  it as a $(4+1)$ scenario.
    
    \end{itemize}
    
    To find the optimal set-up, 
    we generate initial entangled states Haar uniformly. After optimizing parameters in weak measurement and in an auxiliary state, we observe that the maximum average GGM is created in the Scenario 4, i.e., $(4+1)$ with $\sigma_{G}$ being the lowest in this case. In general, we can say, $\langle\mathcal{G}\rangle$  increases with the increase of parties in the initial state and standard deviation behaves oppositely (see Table.~\ref{table:resource_distribution1}). 
    
\begin{figure}
\includegraphics[width=1.0\linewidth]{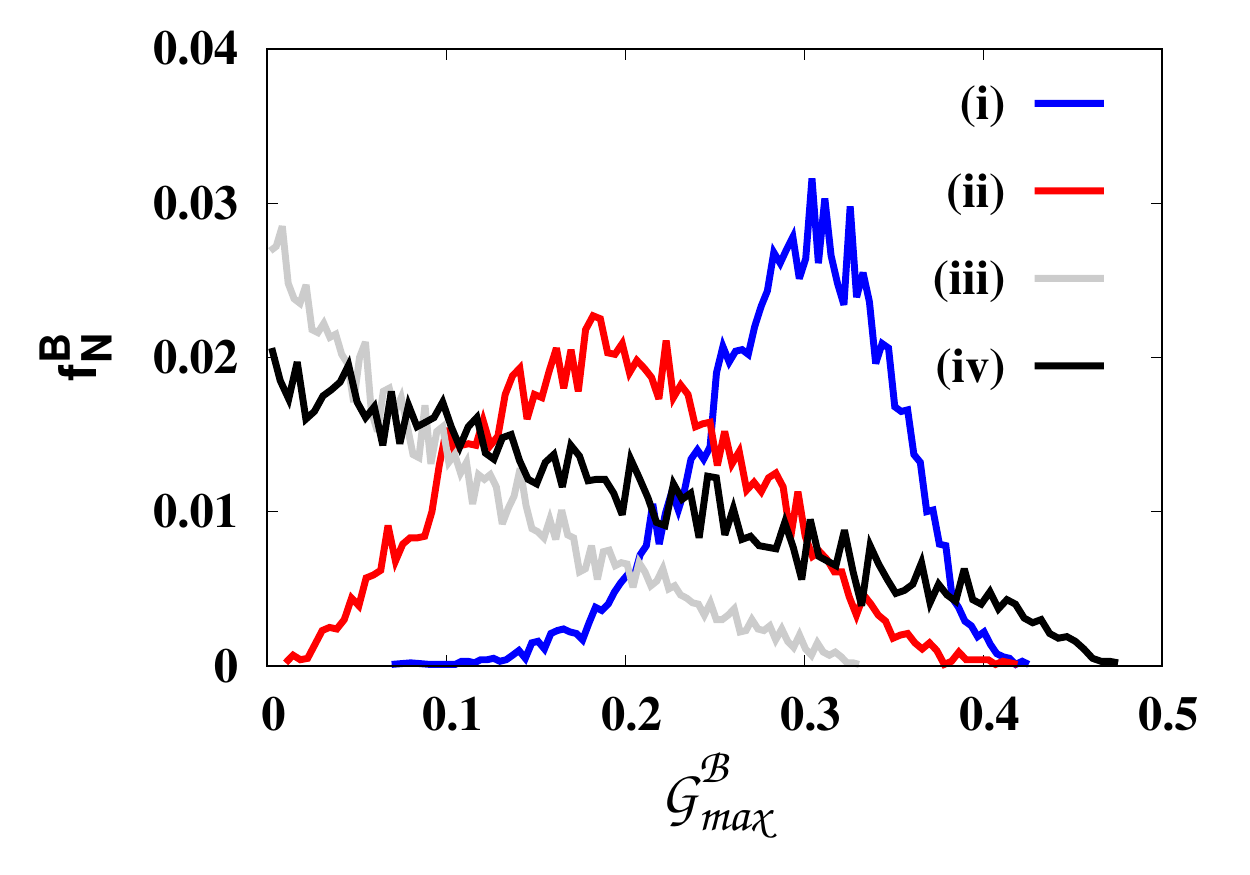}
\caption{(Color online.) Normalized frequency distribution ($f^{B}_{N}$) (ordiante) of $\mathcal{G}_{\max}^{B}$ (abscissa).  \(10^4\) two-qubit, three-qubit W class and GHZ class, and four-qubit initial states are generated Haar uniformly to create five-qubit states after optimizing GGM of the output state for a single outcome, \(M_1^2\).   Here different scenarios are :- (i) $\ket{\Phi}_4$, $\ket{\Phi}_1$, (ii)  $\ket{\Phi}_{3,GHZ}$, $\ket{\Phi}_1$,$\ket{\Phi}_1$, (iii) $\ket{\Phi}_{3,W}$, $\ket{\Phi}_1$,$\ket{\Phi}_1$, (iv) $\ket{\Phi}_2$, $\ket{\Phi}_1$,$\ket{\Phi}_1$,$\ket{\Phi}_1$ with rank-$2$ unsharp measurements in the biased inflation strategy. Both the axes are dimensionless.}
\label{fig:or_biased_hist} 
\end{figure}

    The normalized frequency distributions of GGM of the resulting state, \(f_N^B\) in four different scenarios presented above are depicted in Fig.~\ref{fig:or_biased_hist} by following biased protocol. Clearly, with the increase of the number of parties of the initial entangled state, the distribution shifts towards right while it becomes narrower with the variation of number of parties. Point to note that generating three-qubit  genuinely entangled W-class state is not cost-effective as compared to the situation when  states from the GHZ-class (Scenario 2) or two-qubit random states (Scenario 1) are initially prepared (see Table \ref{table:resource_distribution1}). The entire analysis is carried out when the first element of the rank-$2$ unsharp measurement clicks in the biased state generation protocol. 
    
\begin{figure}
\includegraphics[width=1.0\linewidth]{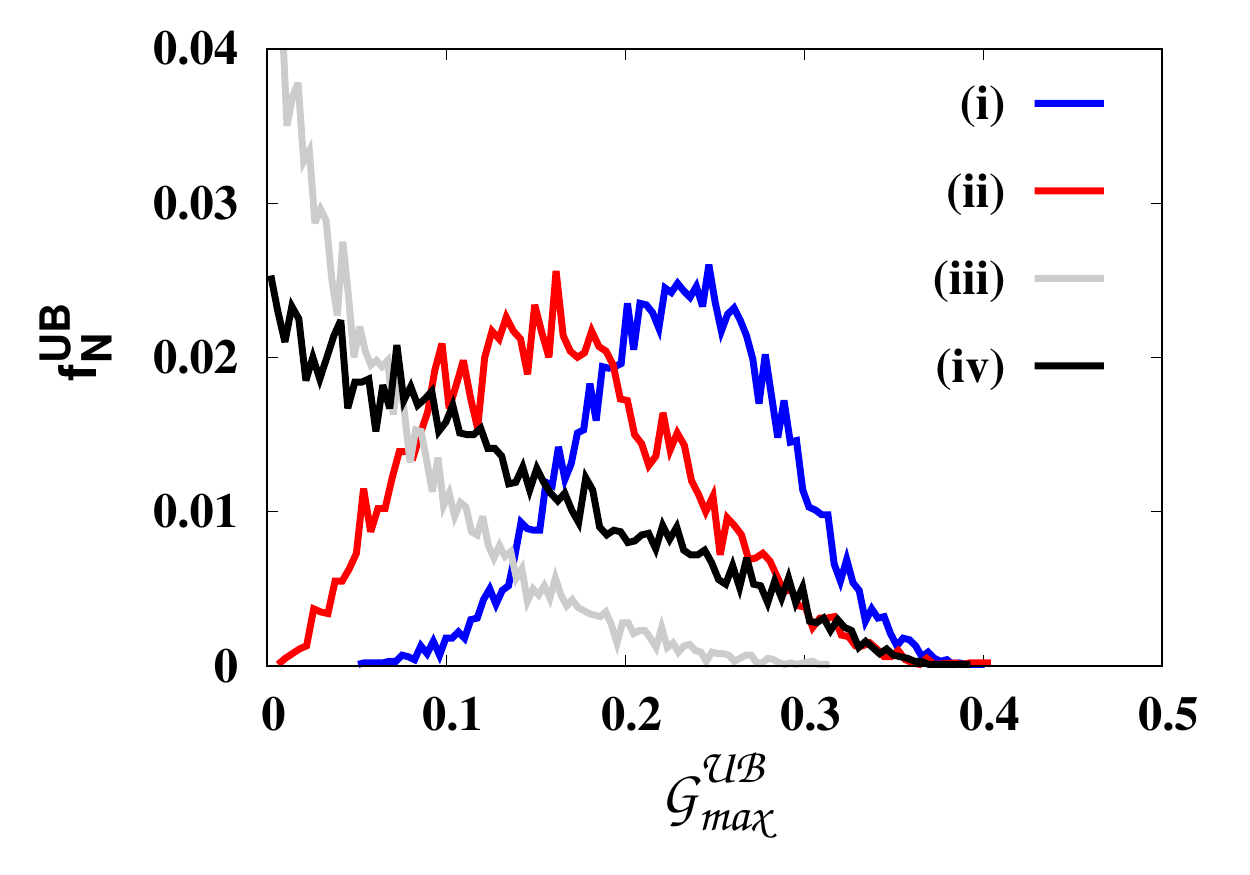}
\caption{(Color online.) Normalized frequency distribution, \(f_N^{UB}\), (ordinate) against $\mathcal{G}_{\max}^{UB}$  (abscissa) for the unbiased inflation process. 
The entire analysis is same as in Fig. \ref{fig:or_biased_hist}. 
Both the axes are dimensionless.}
\label{fig:or_unbiased_hist} 
\end{figure}

    Similar calculations are performed in the case of unbiased inflation protocol and in Fig. ~\ref{fig:or_unbiased_hist},    we plot the normalized frequency distributions of the generated states for different scenarios  (see Table.~\ref{table:resource_distribution2}). In this case too,  different choices of initial resources lead to qualitatively same statistical behavior as we observe for the biased case.


   \begin{table} 
\begin{tabular}{|c|c|c|c|c|c|}
\hline
          \begin{tabular}{c|c}
          No. & Types of unit states \\
          \hline
        (i)  & $\ket{\Phi}_4$, $\ket{\Phi}_1$ \\
          \hline
        (ii) & $\ket{\Phi}_{3,GHZ}$, $\ket{\Phi}_1$,$\ket{\Phi}_1$ \\
          \hline
        (iii)& $\ket{\Phi}_{3,W}$, $\ket{\Phi}_1$,$\ket{\Phi}_1$ \\
          \hline
        (iv) &  $\ket{\Phi}_2$, $\ket{\Phi}_1$,$\ket{\Phi}_1$,$\ket{\Phi}_1$ \\
          
          \end{tabular}
          &
          \begin{tabular}{c|c|c|c}
            $\langle\mathcal{G}\rangle$ & $\sigma_{\mathcal{G}}$ & $\langle\mathcal{T}\rangle$ & $\sigma_{\mathcal{T}}$  \\ 
          \hline 
           $0.2928$  &$0.0514$ & $0.8224$ & $0.1175$   \\
          \hline
         $0.1904$  &$0.0739$  &  $0.5074$ & $0.2010$   \\
         \hline 
           $0.0965$  &$0.0750$ & $0.1676$& $0.1275$ \\
          \hline
         $0.1663$  & $0.1164$ &  $0.4913$& $0.2915$ \\
         \end{tabular}   \\

\hline                      
\end{tabular}
\caption{ Mean and standard deviation of GGM and tangle for the final five-qubit states  corresponding to different choices of  resource distribution with biased strategy of inflation protocol. The computation is carried out when the first element of the unsharp measurement operator with rank-$2$ clicks. The statistical behavior is analyzed with \(10^4\) no. of   haar uniformly generated initial $2$,$3$,$4$-qubit states.  }
\label{table:resource_distribution1}
\end{table}

   \begin{table} 
\begin{tabular}{|c|c|c|c|c|c|}
\hline
          \begin{tabular}{c|c}
          No. & Types of unit states \\
          \hline
        (i)  & $\ket{\Phi}_4$, $\ket{\Phi}_1$ \\
          \hline
        (ii) & $\ket{\Phi}_{3,GHZ}$, $\ket{\Phi}_1$,$\ket{\Phi}_1$ \\
          \hline
        (iii)& $\ket{\Phi}_{3,W}$, $\ket{\Phi}_1$,$\ket{\Phi}_1$ \\
          \hline
        (iv) &  $\ket{\Phi}_2$, $\ket{\Phi}_1$,$\ket{\Phi}_1$,$\ket{\Phi}_1$ \\
          
          \end{tabular}
          &
          \begin{tabular}{c|c|c|c}
            $\langle\mathcal{G}\rangle$ & $\sigma_{\mathcal{G}}$ & $\langle\mathcal{T}\rangle$ & $\sigma_{\mathcal{T}}$  \\ 
          \hline 
           $0.2295$  &$0.0544$ & $0.6131$ & $0.1466$   \\
          \hline
         $0.1619$  &$0.0687$  &  $0.4165$ & $0.1784$   \\
         \hline 
           $0.0633$  &$0561$ & $0.1580$& $0.1504$ \\
          \hline
         $0.1188$  & $0.0861$ &  $0.4711$& $0.3048$ \\
         \end{tabular}   \\

\hline                      
\end{tabular}
\caption{ Mean and standard deviation of GGM and tangle obtained after following unbiased inflation protocol. Notice that unlike Table \ref{tab:unbiased}, here the statistical quantities, mean and the standard deviation are computed by considering \(\mathcal{G}_{\max}^{UB}\). All other specifications are same as in Table \ref{table:resource_distribution1}. }
\label{table:resource_distribution2}
\end{table} 

\section{Conclusion}
\label{sec:conclusion}

In order to create multipartite entangled states, several techniques have been developed which are broadly classified into two categories -  one is deterministic which is typically designed via executing quantum gates while the other ones are measurement-based protocols. Conventional measurement-based protocols start from entangled states with a higher number of parties and build entangled states with fewer parties using local projective measurements. Recently, we have proposed a method based on unsharp measurements to generate multisite entangled states with the aid of the initial two-qubit entangled states and several auxiliary qubits.
\\

This work investigated the potential for producing multiparty entangled states with weak measurements of various characteristics. Specifically, we classify such processes in two distinct ways -  we referred to a biased entanglement inflation protocol where genuine multipartite entanglement of the output state is maximized based on one of the measurement outcomes while in the unbiased picture, average genuine multipartite entanglement is maximized over the parameters for all possible outcomes of the unsharp measurements.

We discovered that the multipartite state with contrasting characters is produced based on measurement features. In particular, we showed that rank-$4$ and rank-$2$ two-qubit weak measurements can only generate the W-class and GHZ-class states, respectively, with two-qubit entangled and single-qubit auxiliary states as resources for the three-qubit generated states. On the other hand, by rank-$3$ measurement operators, it is possible to prepare both the GHZ- and the W-class states.

We observed that depending on the rank of the weak measurement, states with a higher number of qubits can be classified via the measure called persistency. By numerically simulating random initial entangled states, we found that the average genuine multipartite entanglement content decreases with the increase of the rank of the weak measurement operator both in the case of the biased and unbiased protocols. However, the opposite picture emerges for persistency -- it increases with the increase of the rank of measurement operator. It implies that although  low rank measurement operators can create high genuine multpartite entangled states, they are not robust against particle loss. To produce highly entangled multiparty states having a fixed number of parties, we also presented the entangled initial states with an optimal number of parties which should be used as a resource. The overall analysis emphasizes the strength of unsharp measurements in creating multiparty entangled states and hence our investigations  highlight the significance of examining the utility of various types of imperfect measurements in information processing tasks.

\section*{acknwoledgements}

PH, RB, and ASD acknowledge the support from the Interdisciplinary Cyber Physical Systems (ICPS) program of the Department of Science and Technology (DST), India, Grant No.: DST/ICPS/QuST/Theme- 1/2019/23. SM acknowledges the support from Ministry of Science and Technology, Taiwan (Grant No.MOST 110- 2124-M-002-012). We  acknowledge the use of \href{https://github.com/titaschanda/QIClib}{QIClib} -- a modern C++ library for general purpose quantum information processing and quantum computing (\url{https://titaschanda.github.io/QIClib}), and the cluster computing facility at the Harish-Chandra Research Institute. 


\bibliography{ref.bib}

\end{document}